\documentclass[aps,pra,twocolumn,showpacs,superscriptaddress]{revtex4-1}

\usepackage{amsfonts,amssymb,amsmath}
\usepackage{mathtools}
\usepackage[]{graphics,graphicx,epsfig}
\usepackage{amsthm,multirow}
\usepackage{color}
\begin{document}

\title{Optimal entangled coherent states in lossy quantum-enhanced metrology}

\author{Su-Yong \surname{Lee}}
\affiliation{Quantum Physics Technology Directorate, Agency for Defense Development, Daejeon 34186, Korea}

\author{Yong Sup Ihn}
\affiliation{Quantum Physics Technology Directorate, Agency for Defense Development, Daejeon 34186, Korea}

\author{Zaeill Kim}
\affiliation{Quantum Physics Technology Directorate, Agency for Defense Development, Daejeon 34186, Korea}

\date{\today}

\begin{abstract}
We investigate an optimal distance of two components in an entangled coherent state for quantum phase estimation in lossy interferometry. The optimal distance is obtained by an economical point, representing the quantum Fisher information that we can extract per input energy. Maximizing the formula of the quantum Fisher information over an input mean photon number, we show that, as the losses of the interferometer increase, it can be more beneficial to prepare an initially entangled coherent state that is less entangled. This represents that the optimal distance of the two-mode components decreases with more loss in the interferometry. Under the constraint of the input mean photon number, we obtain that the optimal entangled coherent state is more robust than a separable coherent state, even in a high photon loss rate. 
The optimal entangled coherent state preserves quantum advantage over the standard interferometric limit of the separable coherent state.
We also show that the corresponding optimal measurement requires correlation measurement bases.
\end{abstract}

\maketitle

\section{Introduction}

Quantum-enhanced metrology is a feasible quantum protocol under current technology, which is comparable with quantum communication and quantum computation that require perfect security and large-scale architecture, respectively. It can outperform any classical strategies by using entangled or squeezed states as an input signal \cite{giovannetti2004}, due to their nonclassical behaviors.
For example, the $N00N$ state exhibits $N$-time oscillation in a single run whereas the coherent state exhibits single-time oscillation in the same run \cite{dowling2008}. 
In comparison to the coherent state which represents the number-phase uncertainty with equality, the squeezed state reduces the uncertainty of phase while it enlarges that of number (intensity) \cite{dowling2008}. It is known that, for the standard quantum state, one utilizes the coherent state that is a minimum uncertainty state and imitates the oscillatory behavior of a classical harmonic oscillator. 

We focus on the two-mode input state for quantum phase estimation in lossy interferometry. In the lossless scenario, the $N00N$ state \cite{dowling2008} and
$N00N$-type states \cite{Gerry01,Joo11, Zhang13, LLNK15, Knott2016, LLLN16} demonstrate the Heisenberg limit(HL) which is the fundamental limit by quantum mechanics, representing a scaling of $1/N$.
The state provides enhancement of $\sqrt{N}$ by entanglement in precision whereas coherent state represents a scaling of $1/\sqrt{N}$, i.e., 
the shot-noise limit (SNL) or the standard quantum limit (SQL). $N$ is the input photon number, which can be replaced by the mean photon number of the input state.
In the lossy scenario, however, the $N00N$ state and $N00N$-type states are fragile against photon loss, such that people studied a class of path-entangled photon states \cite{Huver08}, Holland-Burnett states \cite{HB, Datta}, and general pure states with definite photon number $N$ \cite{Dorner, Rafal} in order to be resilient to photon loss.
Although a pure state with definite photon number $N=2$ was optimized experimentally to show one of the best states in the lossy quantum-enhanced metrology\cite{Kac}, we could not find any relation between two components ($|\rangle_a$ and $|\rangle_b$) in the pure state. Here we raise a question: is there any specific relation between two components of a two-mode state in lossy quantum-enhanced metrology? For a specific relation, we consider an optimal distance of two components in a two-mode state which varies with photon loss in an interferometry. 

%Experimentally, the pure state with definite photon number $N=2$ showed better performance than the other states in the presence of photon loss \cite{Kac}, while optimizing the coefficients of the number bases, i.e., $|2\rangle_a|0\rangle_b$, $|1\rangle_a|1\rangle_b$, and $|0\rangle_a|2\rangle_b$, for different photon loss rates.In spite of the result, we do not know what is the optimal distance between two components in the two-mode states for quantum phase estimation in lossy interferometry. Here we raise a question as to how the optimal distance of the two-mode state components varies with photon loss in the interferometry.
It is related to initially prepare an optimal input state against photon loss in order to avoid additional operations in the measurement stage.  
The optimal input state preserves quantum advantage in the presence of photon loss, while beating the standard interferometric limit (SIL) of a coherent state. In the presence of photon loss, the boundary of the coherent state becomes readjusted as the SIL \cite{Rafal}.
Here we inject a coherent input state ($|\sqrt{2}\alpha\rangle_a |0\rangle_b$) into an interferometer.  After a $50:50$ beam splitter, the input coherent state is transformed into a separable coherent state ($|\alpha\rangle_a |\alpha\rangle_b$) which is compared with our optimal entangled coherent state in the presence of photon loss. Thus, in our paper, the separable coherent state plays roles of the SQL in the absence of photon loss as well as the SIL in the presence of photon loss.

% In this work, we propose optimal entangled coherent states $|\alpha\rangle_a|\beta\rangle_b+|\beta\rangle_a|\alpha\rangle_b$ for quantum phase estimation in the presence of photon loss.  
In general, an arbitrary two-mode entangled state can be represented by $|\psi_1\rangle_a|\psi_2\rangle_b+|\psi_2\rangle_a|\psi_1\rangle_b$,   
where $|\psi_{1(2)} \rangle$ is an arbitrary single-mode state. It is required to specify them in order to observe the distance between the two components $\psi_1$ and $\psi_2$. In continuous variable systems, coherent states are more feasible than the other states, e.g., the squeezed vacuum state and squeezed coherent state. 
 Using the coherent states, we propose optimal entangled coherent states $|\alpha\rangle_a|\beta\rangle_b+|\beta\rangle_a|\alpha\rangle_b$ for quantum phase estimation in the presence of photon loss. 
The entangled coherent state, which is implementable in a laboratory \cite{cat}, is a good candidate to control the optimal distance between the two components in the two-mode state. 
In Fig. 1, we simply put the photon-loss process after the phase-shifting process, due to the commutativity between the two processes \cite{Rafal,Oh17}.
We investigate the optimal distance of $\alpha$ and $\beta$ for quantum phase estimation with different photon loss rates $R$. The optimal distance is obtained by an economical point, representing the maximum information that we can extract per input energy.
Using the formula of the quantum Fisher information (QFI) over the input mean photon number, we show how the optimal distance of $\alpha$ and $\beta$ varies with the photon loss rate in the interferometry.  The result is also explained with the degree of entanglement (DOE) for the optimal entangled coherent state. 
In the constraint of the input mean photon number, the optimal entangled coherent state is compared with a separable coherent state in the presence of photon loss. 
Furthermore, we derive the optimal measurement for the quantum Fisher information.

This paper is organized as follows. We propose a generation scheme of an entangled coherent state and analyze its degree of entanglement.
Then we investigate optimal distance of the coherent state components $\alpha$ and $\beta$ with photon loss rate, and compare the optimal entangled coherent state with a separable coherent state. It is also analyzed by the output state entanglement.
Next, we discuss the corresponding optimal measurement. Finally, we summarize our results and discuss some issues.

\section{State generation and degree of entanglement}

\begin{figure}
\centerline{\scalebox{0.35}{\includegraphics[angle=0]{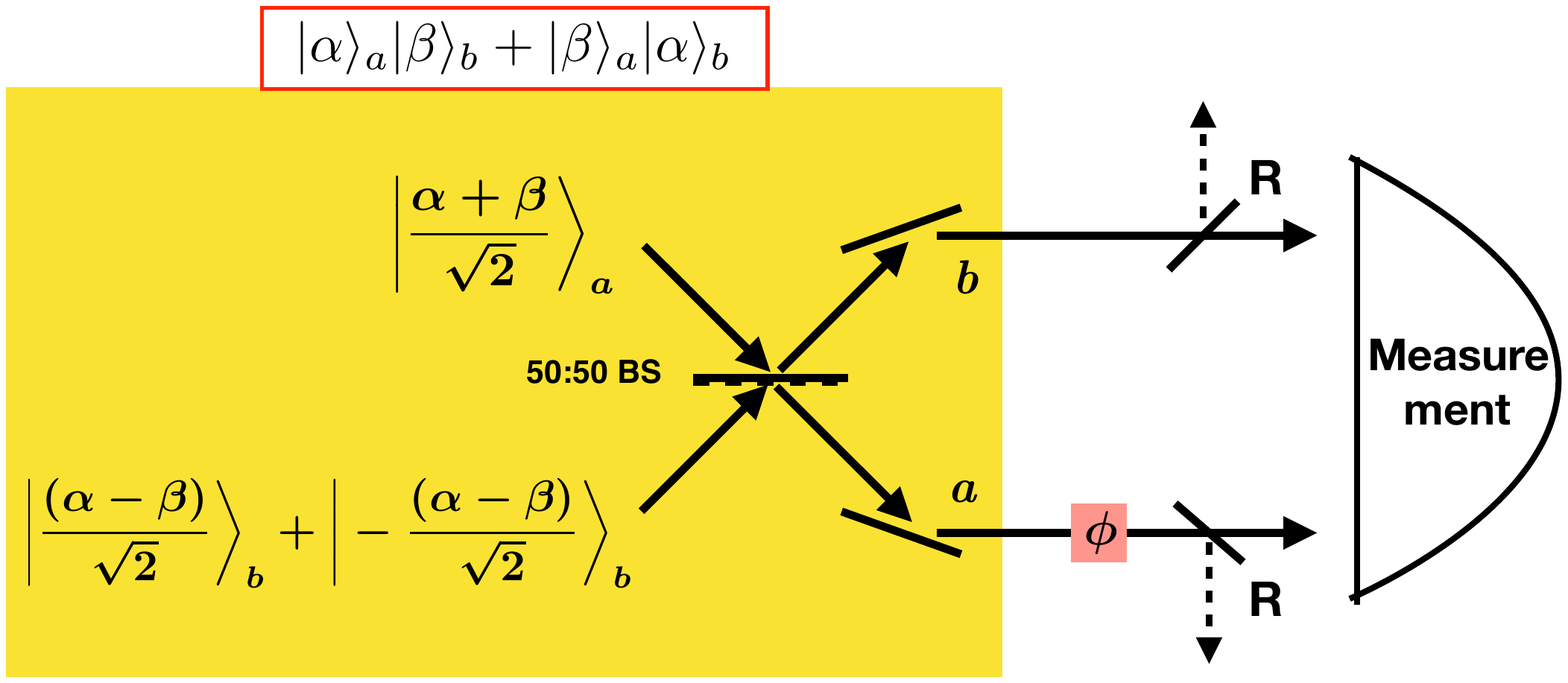}}}
\vspace{-1.6in}
\caption{Entangled coherent state for quantum phase estimation in the presence of photon loss. $R$ is a photon loss rate.
$\phi$ represents  a phase shifter, $\exp(i\phi\hat{n})$. The input entangled coherent state $|\alpha\rangle_a|\beta\rangle_b+|\beta\rangle_a|\alpha\rangle_b$ is prepared after a $50:50$ beam splitter.}
\label{fig:fig1}
\end{figure}

An input entangled coherent state can be produced by impinging a coherent state and an even cat state into a $50:50$ beam splitter which takes
the transformation of $\hat{a}^{\dag}\rightarrow \frac{1}{\sqrt{2}}(\hat{a}^{\dag}+\hat{b}^{\dag})$ and
$\hat{b}^{\dag}\rightarrow \frac{1}{\sqrt{2}}(\hat{b}^{\dag}-\hat{a}^{\dag})$.
In Fig. 1, we inject a coherent state $|\frac{\alpha+\beta}{\sqrt{2}}\rangle_a$ and an even cat state 
$|\frac{(\alpha-\beta)}{\sqrt{2}}\rangle_b+|-\frac{(\alpha-\beta)}{\sqrt{2}}\rangle_b$ into the $50:50$ beam splitter. Then, we produce an entangled coherent state $|\alpha\rangle_a|\beta\rangle_b+|\beta\rangle_a|\alpha\rangle_b$, 
where a single-mode coherent state is given by $|\alpha\rangle=e^{-|\alpha|^2/2}\sum^{\infty}_{n=0}\frac{\alpha^n}{\sqrt{n!}}|n\rangle$.
Assuming that $\alpha$ and $\beta$ are real variables, we can transform the entangled coherent state into
\begin{eqnarray}
&&\frac{1}{\sqrt{N_T}}(|\alpha\rangle_a|\beta\rangle_b+|\beta\rangle_a|\alpha\rangle_b)\\
&&=\frac{1}{2\sqrt{N_T}}(\langle A_+|A_+\rangle|\bar{A}_+\rangle_a|\bar{A}_+\rangle_b-\langle A_-|A_-\rangle|\bar{A}_-\rangle_a|\bar{A}_-\rangle_b),\nonumber
\end{eqnarray}
where $|A_{\pm}\rangle=|\alpha\rangle \pm |\beta\rangle$, $|\bar{A}_{\pm}\rangle=\frac{1}{\sqrt{\langle A_{\pm}|A_{\pm}\rangle}}|A_{\pm}\rangle$, 
$\langle A_-|A_+\rangle=0$, and $N_T=2(1+\exp[-(\beta-\alpha)^2])$.
Using the von Neumann entropy \cite{ent} which is computed as $DOE\equiv-Tr[\rho_a\log_2\rho_a]=-Tr[\rho_b\log_2\rho_b]=-\sum_i\lambda_i\log_2\lambda_i$ ($\lambda_i$ denotes the eigenvalues of $\rho_a$ or $\rho_b$) for a pure bipartite state, 
we derive the DOE as
\begin{eqnarray}
\text{DOE}=-\sum_{k=+,-}\frac{\langle A_k|A_k\rangle^2}{4N_T}\log_2[\frac{\langle A_k|A_k\rangle^2}{4N_T}],
\end{eqnarray}
where $\langle A_{\pm}|A_{\pm}\rangle=1\pm \exp[-(\beta-\alpha)^2/2]$.
In Fig. 2, we show that the degree of entanglement increases with $|\alpha-\beta|$.  Since we assumed that $\alpha$ and $\beta$ are real values, we can take 
$|\alpha-\beta|$ as the distance between the two components $\alpha$ and $\beta$. This implies that the two different states 
$|2\alpha\rangle_a|0\rangle_b+|0\rangle_a|2\alpha\rangle_b$ and $|\alpha\rangle_a|-\alpha\rangle_b+|-\alpha\rangle_a|\alpha\rangle_b$ have the same degree of entanglement which counted their normalization factors.

\begin{figure}
\centerline{\scalebox{0.6}{\includegraphics[angle=0]{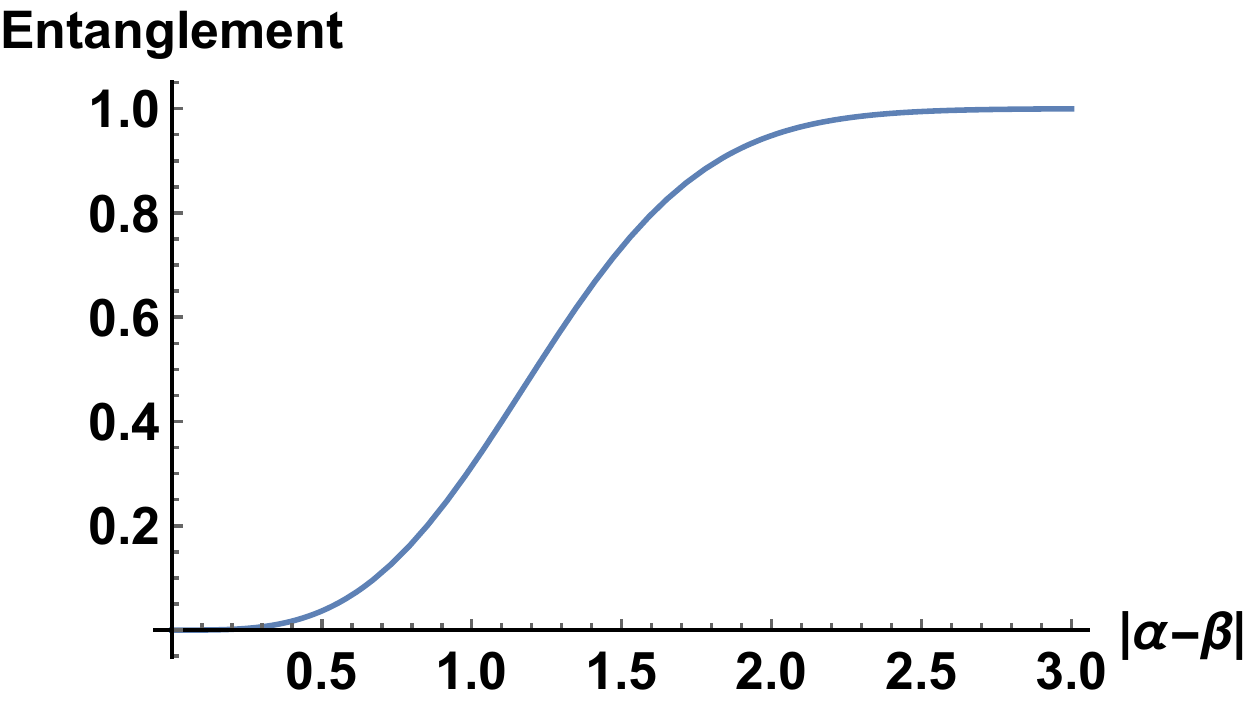}}}
\vspace{0in}
\caption{Degree of entanglement for the entangled coherent state $|\alpha\rangle_a|\beta\rangle_b+|\beta\rangle_a|\alpha\rangle_b$, 
as a function of $|\alpha-\beta|$. }
\label{fig:fig2}
\end{figure}

Experimentally, in superconducting circuit-QED systems,  an even cat state $|\frac{(\alpha-\beta)}{\sqrt{2}}\rangle+|-\frac{(\alpha-\beta)}{\sqrt{2}}\rangle$ was prepared with the fidelity more than $99\%$, and an entangled coherent state $|\alpha\rangle_a|\alpha\rangle_b+|-\alpha\rangle_a|-\alpha\rangle_b$ was prepared with $81\%$ fidelity at $\alpha=1.92$ \cite{cat}. It is expected that the $|\alpha\rangle_a|\beta\rangle_b+|\beta\rangle_a|\alpha\rangle_b$ state can be also prepared with the fidelity more than $80\%$ in the superconducting circuit-QED.

\section{Optimal distance of two components ($\alpha$ and $\beta$) in an entangled coherent state}
Using the entangled coherent state $|\alpha\rangle_a|\beta\rangle_b+|\beta\rangle_a|\alpha\rangle_a$, we study quantum phase estimation in lossy interferometry. It can be quantified by means of the quantum Fisher information that is the optimal measure of how well we can detect small changes in a parameter, which is the maximized classical Fisher information (CFI) with positive-operator valued measures. 
From an information-theoretic point of view, the inverse of the QFI can determine the ultimate precision limit of quantum phase estimation \cite{BK}.
This implies that the more the QFI is, the better the precision limit of a phase is.
The QFI is defined as $F_Q=Tr[\rho_{\phi}\hat{L}_{\phi}^2]$, where $\rho_{\phi}$ contains the phase information of $\phi$ and $\hat{L}_{\phi}$ is the symmetric logarithmic derivative (SLD) operator \cite{BK}.
Using an output state $\rho_{\phi}=\sum_n \lambda_n|\lambda_n\rangle\langle\lambda_n|$, we can obtain the QFI by a formula of $F_Q=4\sum_n\lambda_n f_n-\sum_{n\neq m}\frac{8\lambda_n\lambda_m}{\lambda_n + \lambda_m}|\langle \lambda^{'}_n|\lambda_m\rangle|^2$, where $f_n=\langle\lambda^{'}_n| \lambda^{'}_n\rangle-|\langle\lambda^{'}_n|\lambda_n\rangle|^2$ and 
$| \lambda^{'}_n\rangle=\frac{\partial |\lambda_n\rangle}{\partial \phi}$.

\begin{figure}
\centerline{\scalebox{0.6}{\includegraphics[angle=0]{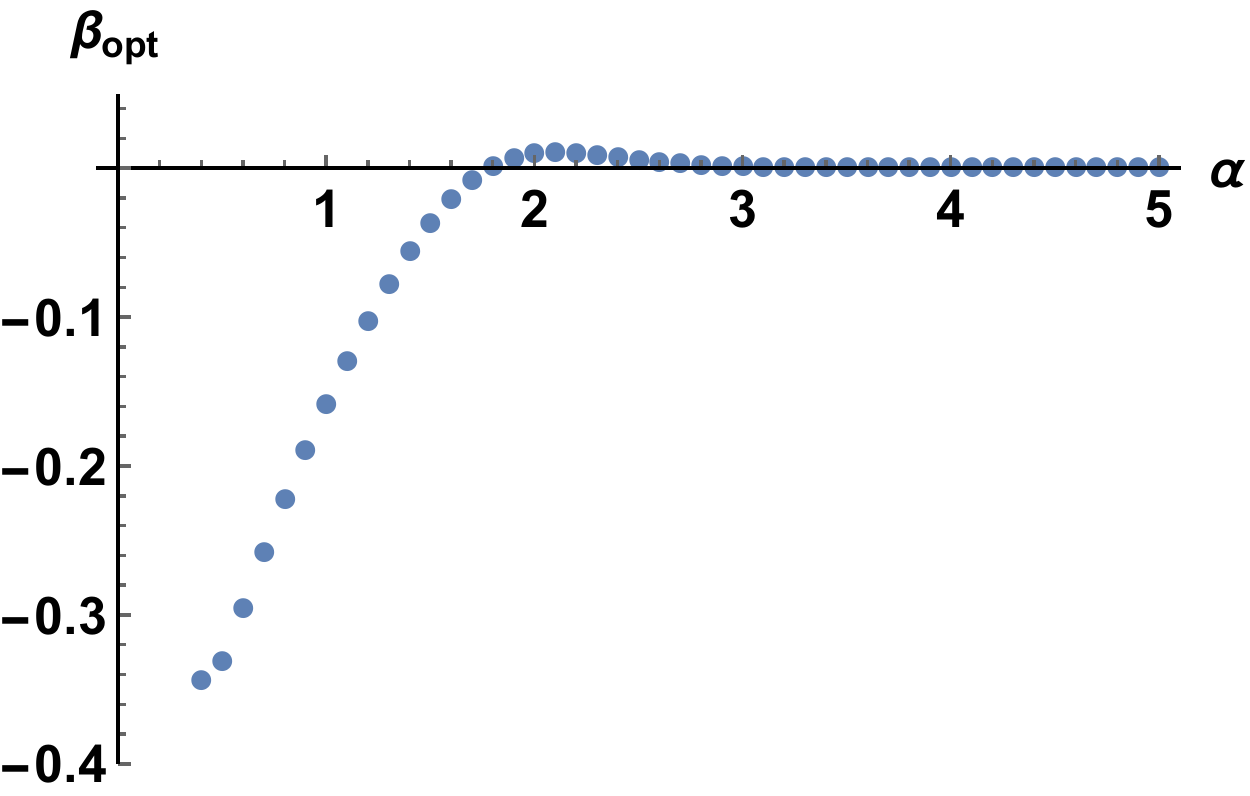}}}
\vspace{0in}
\caption{Economical point of quantum phase estimation using the state $|\alpha\rangle_a|\beta\rangle_b+|\beta\rangle_a|\alpha\rangle_b$ without photon loss.
Given a  value of $\alpha$, we find the optimal value of $\beta$ achieving the maximum value of $F_Q/\langle \hat{n}_a\rangle$, the quantum Fisher information over the input mean photon number in mode $a$. At $\alpha <0.4$, $\beta_{opt}=-\alpha$ which were not shown here. 
The optimal value of $\beta$ is numerically obtained under the constraint of $-\alpha \leq \beta < \alpha$. 
}
\label{fig:fig3}
\end{figure}

Here we are interested in the quantum phase estimation for the cost-effective use of the input entangled coherent state.
This means that we want to acquire more information of a phase parameter per input energy.
For a measure of the cost-effectiveness, we define an economical point as 
\begin{equation}
\text{Eco}(R, \alpha, \beta)\equiv \max (\frac{F_Q}{\langle \hat{n}_a\rangle}),
\end{equation}
where $F_Q$ is the QFI of the output state in Fig. 1 and $\langle \hat{n}_a\rangle$ is the input mean photon number in mode $a$ after the $50:50$ beam splitter. Since the total input mean photon number is satisfied with the condition $\langle \hat{n}_a+\hat{n}_b\rangle=2\langle\hat{n}_a\rangle$, it is enough for us to consider either of the two modes as an input energy. 
Given a  value of $\alpha$, we find the optimal value of $\beta$ in the range of $-\alpha\leq \beta< \alpha$ to maximize $F_Q/\langle \hat{n}_a\rangle$.

\subsection{Equal losses in both arms}
After experiencing a phase shifting operation and photon loss, the output state is given by
%\begin{widetext}
\begin{eqnarray}
\rho_{\phi}&=&\frac{1}{2[1+e^{-(\beta-\alpha)^2}]}\nonumber\\
&\times&\bigg[ |\sqrt{T}\alpha e^{i\phi}\rangle_a\langle \sqrt{T}\alpha e^{i\phi}|\otimes |\sqrt{T}\beta \rangle_b\langle \sqrt{T}\beta |\nonumber\\
&&+ |\sqrt{T}\beta e^{i\phi}\rangle_a\langle \sqrt{T}\beta e^{i\phi}|\otimes |\sqrt{T}\alpha \rangle_b\langle \sqrt{T}\alpha | \nonumber\\
&& +e^{-R(\beta-\alpha)^2}\bigg( |\sqrt{T}\alpha e^{i\phi}\rangle_a\langle \sqrt{T}\beta e^{i\phi}|\otimes |\sqrt{T}\beta \rangle_b\langle \sqrt{T}\alpha|\nonumber\\
&&+ |\sqrt{T}\beta e^{i\phi}\rangle_a\langle \sqrt{T}\alpha e^{i\phi}|\otimes |\sqrt{T}\alpha \rangle_b\langle \sqrt{T}\beta|\bigg) \bigg],
\end{eqnarray}
%\end{widetext}
where $T=1-R$. The corresponding quantum Fisher information is derived as
\begin{eqnarray}
F_Q&=&4\bigg[ \sum_{k=+,-}\lambda_k(\langle \lambda^{'}_k|\lambda^{'}_k\rangle -|\langle \lambda^{'}_k|\lambda_k\rangle|^2)\nonumber\\
&&-\frac{2\lambda_+\lambda_-}{\lambda_+ +\lambda_-}(|\langle \lambda^{'}_+|\lambda_-\rangle|^2+|\langle \lambda^{'}_-|\lambda_+\rangle|^2) \bigg],
\end{eqnarray}
where
\begin{eqnarray}
\langle \lambda^{'}_{\pm}|\lambda^{'}_{\pm}\rangle&=&\frac{T}{2[1\pm e^{-T(\beta-\alpha)^2}]}
\bigg[ \alpha^2(T\alpha^2+1)+\beta^2(T\beta^2+1)\nonumber\\
&&\pm 2\alpha\beta(T\alpha\beta+1)e^{-T(\beta-\alpha)^2}\bigg],\nonumber\\
\langle \lambda^{'}_{\pm}|\lambda_{\pm}\rangle&=&\frac{-iT}{2[1\pm e^{-T(\beta-\alpha)^2}]}
\bigg[ \alpha^2+\beta^2\pm 2\alpha\beta e^{-T(\beta-\alpha)^2}\bigg],\nonumber\\
\langle \lambda^{'}_{+}|\lambda_{-}\rangle&=& \langle \lambda^{'}_{-}|\lambda_{+}\rangle=
\frac{-iT(\alpha^2-\beta^2)}{2\sqrt{1-e^{-2T(\beta-\alpha)^2}}},\nonumber\\
\lambda_{\pm}&=&\frac{(1\pm e^{-R(\beta-\alpha)^2})}{2(1+e^{-(\beta-\alpha)^2})}(1\pm e^{-T(\beta-\alpha)^2}).
\end{eqnarray}

\begin{figure}
\centerline{\scalebox{0.31}{\includegraphics[angle=0]{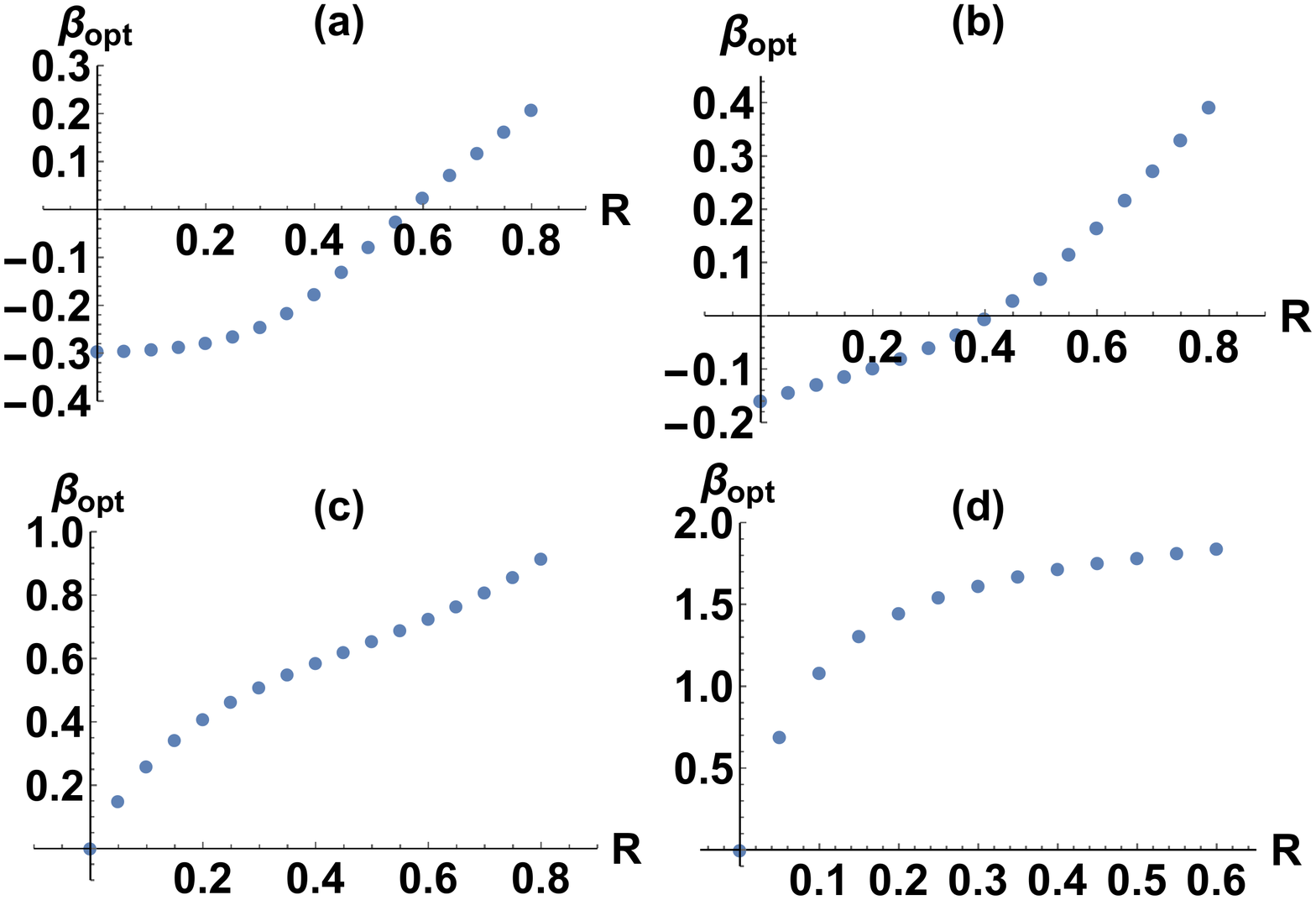}}}
\vspace{-0.3in}
\caption{(Photon loss in both arms) Economical point of quantum phase estimation using the state $|\alpha\rangle_a|\beta\rangle_b+|\beta\rangle_a|\alpha\rangle_b$ with photon loss rate $R$.
Give a  value of $\alpha$, we find the optimal value of $\beta$ achieving the maximum value of $F_Q/\langle \hat{n}_a\rangle$, the quantum Fisher information over the input mean photon number in mode $a$, depending on the photon loss rate $R$. (a) $\alpha=0.6$, (b) $\alpha=1$, (c) $\alpha=1.8$, and (d) $\alpha=3$.
We consider the constraint of $-\alpha \leq \beta < \alpha$. }
\label{fig:fig4}
\end{figure}

In a lossless condition $(R=0)$, given a value of $\alpha$, the optimal value of $\beta$ is negative at $\alpha<1.8$ but it increases with $\alpha$ and approaches zero at $\alpha\geq 1.8$, as shown in Fig. 3.  At $\alpha <0.4$, the optimal value of $\beta$ is given as $-\alpha$ but we did not show it in Fig. 3.
For small values of $\alpha$, the optimal entangled coherent state is close to the state $|\alpha\rangle_a|-\alpha\rangle_b+|-\alpha\rangle_a|\alpha\rangle_b$. For large values of $\alpha$, the optimal entangled coherent state becomes the state $|\alpha\rangle_a|0\rangle_b+|0\rangle_a|\alpha\rangle_b$.
We note that the former state is close to a scaling of the SQL and the latter state approaches a scaling of the HL. With increasing value of $\alpha$, the ultimate precision limit of the optimal entangled coherent state proceeds from the SQL to the HL, while the optimal distance of $|\alpha-\beta|$ varies from $|2\alpha|$ to $|\alpha|$. 
%We note that the phase sensitivity by the QFI at $\beta= -\alpha$ is a scaling of the SQL even if the input two-mode coherent state can exhibit the maximal entanglement at $\beta=-\alpha$.

In a lossy condition $(R\neq 0)$, given a value of $\alpha$, the optimal value of $\beta$ increases with the photon loss rate $R$. This represents that the optimal distance 
of $|\alpha-\beta|$ decreases with the photon loss rate. In Fig. 4, we show that the optimal distance is getting smaller (larger) with the increasing (decreasing) photon loss rate. This means that initially we need to prepare less (more) entangled coherent states under the increasing (decreasing) photon loss rate. At $\alpha=0.6$, the optimal value of $\beta$ increases approximately from $-0.30$ to $0.21$ with the photon loss rate $R$.
At $\alpha=1$, it moves from $-0.16$ to $0.39$. At $\alpha=1.8$, it starts to move from $0$ to $0.92$. Then, at $\alpha=3$, it shifts from $0$ to $1.84$.
We note that, in Fig. 4 (a)-(c), the optimal value of $\beta$ can be any value of $\alpha$ under $R>0.8$ so that the optimal entangled coherent state can be a separable state $|\alpha\rangle_a|\alpha\rangle_b$. It is also applied for the case of Fig. 4 (d) under $R>0.6$. 

\begin{figure}
\centerline{\scalebox{0.31}{\includegraphics[angle=0]{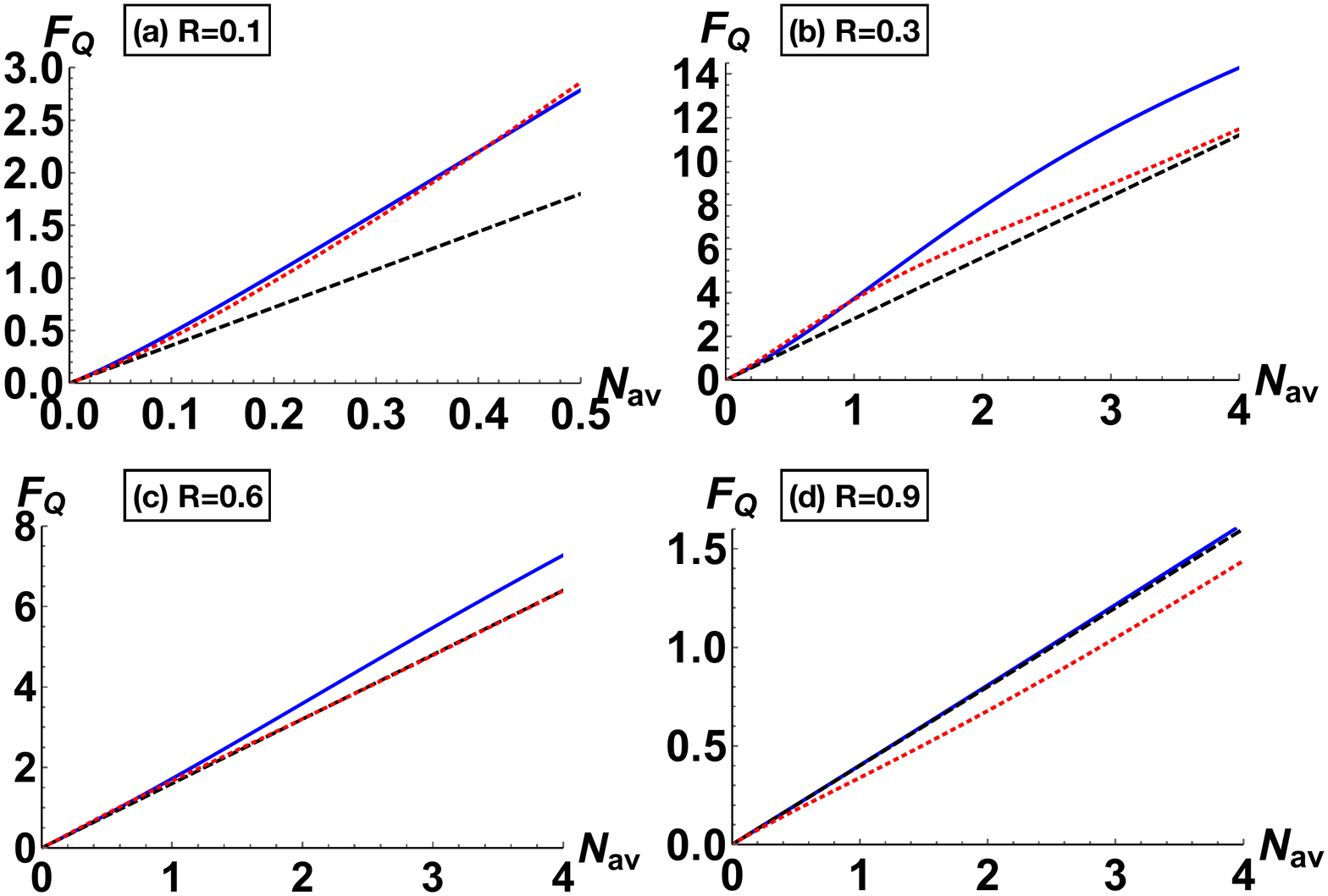}}}
\vspace{-0.3in}
\caption{(Photon loss in both arms) Comparison between the state $|\alpha\rangle_a|\beta\rangle_b+|\beta\rangle_a|\alpha\rangle_b$ and a separable coherent state $|\alpha\rangle_a|\alpha\rangle_b$ for the quantum Fisher information with $N_{\text{av}}=\langle\hat{n}_a\rangle$, in the constraint of input mean photon number. $R$ is the photon loss rate.
Black dashed line ($|\alpha\rangle_a|\alpha\rangle_b$), red dotted curve ($|\alpha\rangle_a|0\rangle_b+|0\rangle_a|\alpha\rangle_b$), 
and blue solid curve ($|\alpha\rangle_a|\beta\rangle_b+|\beta\rangle_a|\alpha\rangle_b$).
(a) $\beta=-0.2\alpha$, (b) $\beta=0.3\alpha$, (c) $\beta=0.5\alpha$, and (d) $\beta=0.7\alpha$.
}
\label{fig:fig5}
\end{figure}

In Fig. 5, the results of the entangled coherent state $|\alpha\rangle_a|\beta\rangle_b+|\beta\rangle_a|\alpha\rangle_b$ can be compared with those of a separable coherent state $|\alpha\rangle_a|\alpha\rangle_b$. Given a mean photon number $|\alpha|^2$ for the coherent state, the quantum Fisher information is derived as $4(1-R)|\alpha|^2$ in the presence of photon loss rate $R$. Under the constraint of the input mean photon number, the entangled coherent state outperforms the coherent state by means of the QFI. In a lossless condition ($R=0$), the entangled coherent state with $\beta\sim -\alpha$ exhibits higher QFI than the coherent state at the small mean photon number, whereas the entangled coherent state with $\beta=0$ exhibits higher QFI than the coherent state at the large mean photon number. In a lossy condition ($R\neq 0$), the entangled coherent state with $\beta=\gamma \alpha$ shows higher QFI than the coherent state for different photon loss rates $R$, where the ratio $\gamma$ increases from $-1$ to less than $1$ with the increasing $R$.

\subsection{Losses in one arm}

We can consider the same scenario under a photon loss only in one arm of the interferometry. Assuming the photon loss only in the mode $a$ of Fig. 1, we derive the output state as
%\begin{widetext}
\begin{eqnarray}
\rho_{\phi}&=&\frac{1}{2[1+e^{-(\beta-\alpha)^2}]}
\bigg[ |\sqrt{T}\alpha e^{i\phi}\rangle_a\langle \sqrt{T}\alpha e^{i\phi}|\otimes |\beta \rangle_b\langle \beta |\nonumber\\
&&+ |\sqrt{T}\beta e^{i\phi}\rangle_a\langle \sqrt{T}\beta e^{i\phi}|\otimes |\alpha \rangle_b\langle \alpha | \nonumber\\
&& +e^{-R(\beta-\alpha)^2}\bigg( |\sqrt{T}\alpha e^{i\phi}\rangle_a\langle \sqrt{T}\beta e^{i\phi}|\otimes |\beta \rangle_b\langle \alpha|\nonumber\\
&&+ |\sqrt{T}\beta e^{i\phi}\rangle_a\langle \sqrt{T}\alpha e^{i\phi}|\otimes |\alpha \rangle_b\langle \beta|\bigg) \bigg],
\end{eqnarray}
%\end{widetext}
where $T=1-R$. The QFI of the Eq. (5) is calculated with 
\begin{eqnarray}
\langle \lambda^{'}_{\pm}|\lambda^{'}_{\pm}\rangle
&=&\frac{T}{2[1\pm e^{-\frac{1}{2}(1+T)(\beta-\alpha)^2}]}
\bigg[ \alpha^2(T\alpha^2+1) +\beta^2\nonumber\\
&\times&(T\beta^2+1)\pm 2\alpha\beta(T\alpha\beta+1)e^{-\frac{1}{2}(1+T)(\beta-\alpha)^2}\bigg],\nonumber\\
\langle \lambda^{'}_{\pm}|\lambda_{\pm}\rangle&=&\frac{-iT\bigg[ \alpha^2+\beta^2\pm 2\alpha\beta e^{-\frac{1}{2}(1+T)(\beta-\alpha)^2}\bigg]}{2[1\pm e^{-\frac{1}{2}(1+T)(\beta-\alpha)^2}]}
,\nonumber\\
\langle \lambda^{'}_{+}|\lambda_{-}\rangle&=& \langle \lambda^{'}_{-}|\lambda_{+}\rangle=
\frac{iT(\alpha^2-\beta^2)}{2\sqrt{1-e^{-(1+T)(\beta-\alpha)^2}}},\nonumber\\
\lambda_{\pm}&=&\frac{(1\pm e^{-\frac{R}{2}(\beta-\alpha)^2})}{2(1+e^{-(\beta-\alpha)^2})}(1\pm e^{-\frac{1}{2}(1+T)(\beta-\alpha)^2}).
\end{eqnarray}
It exhibits similar behavior to Fig. 4. Given a value of $\alpha$, the optimal value of $\beta$ increases with the photon loss rate $R$, such that the optimal distance of $|\alpha-\beta|$ decreases with the photon loss rate, as shown in Fig. 6. Using the QFI, in Fig. 7, we show that the entangled coherent state exhibits higher QFI than the coherent state with increasing $R$. 

\begin{figure}
\centerline{\scalebox{0.31}{\includegraphics[angle=0]{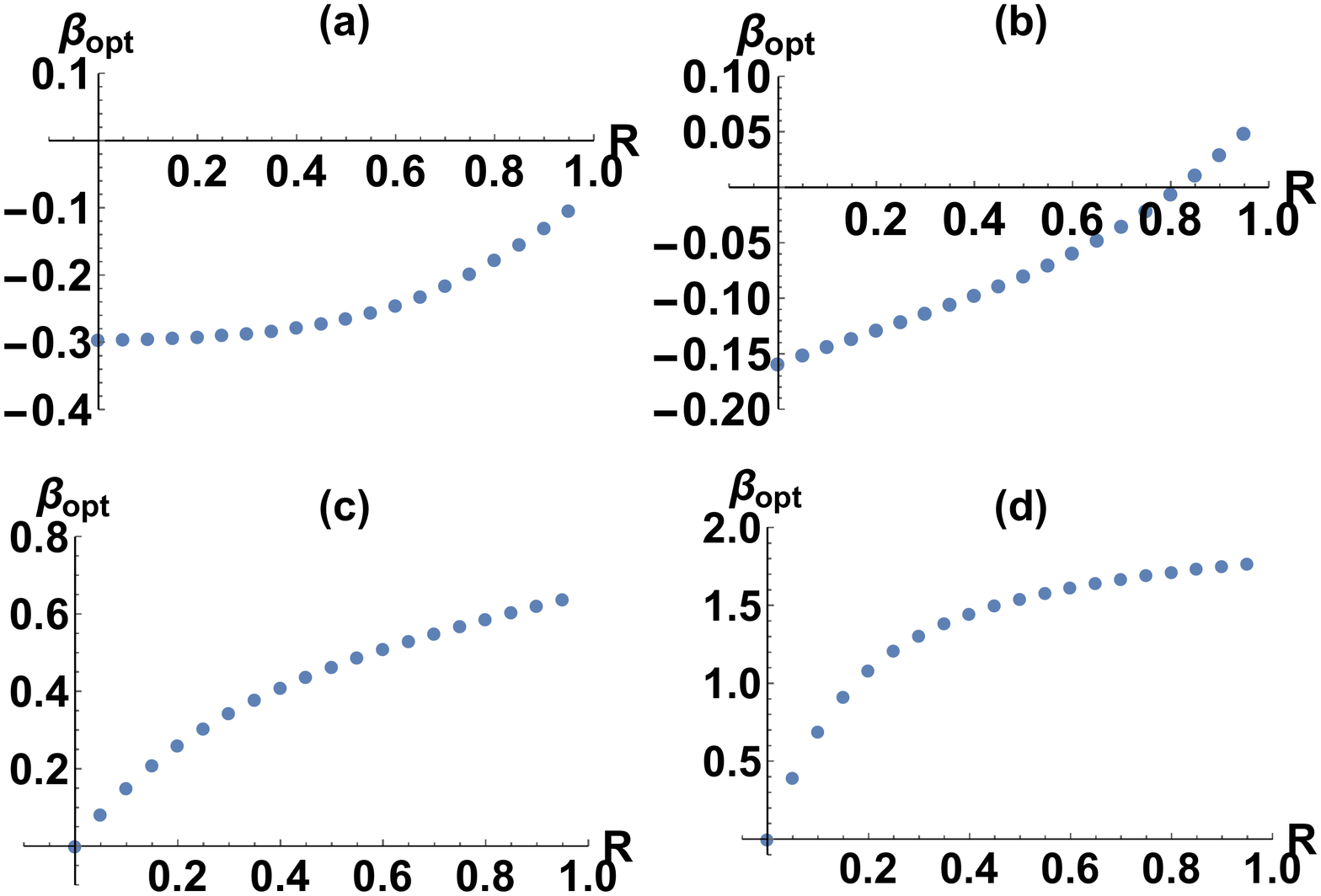}}}
\vspace{-0.3in}
\caption{(Photon loss in one arm) Economical point of quantum phase estimation using the state $|\alpha\rangle_a|\beta\rangle_b+|\beta\rangle_a|\alpha\rangle_b$ with photon loss rate $R$.
Give a  value of $\alpha$, we find the optimal value of $\beta$ achieving the maximum value of $F_Q/\langle \hat{n}_a\rangle$, the quantum Fisher information over the input mean photon number in mode $a$, depending on the photon loss rate $R$. (a) $\alpha=0.6$, (b) $\alpha=1$, (c) $\alpha=1.8$, and (d) $\alpha=3$.
We consider the constraint of $-\alpha \leq \beta < \alpha$. }
\label{fig:fig6}
\end{figure}

\begin{figure}
\centerline{\scalebox{0.31}{\includegraphics[angle=0]{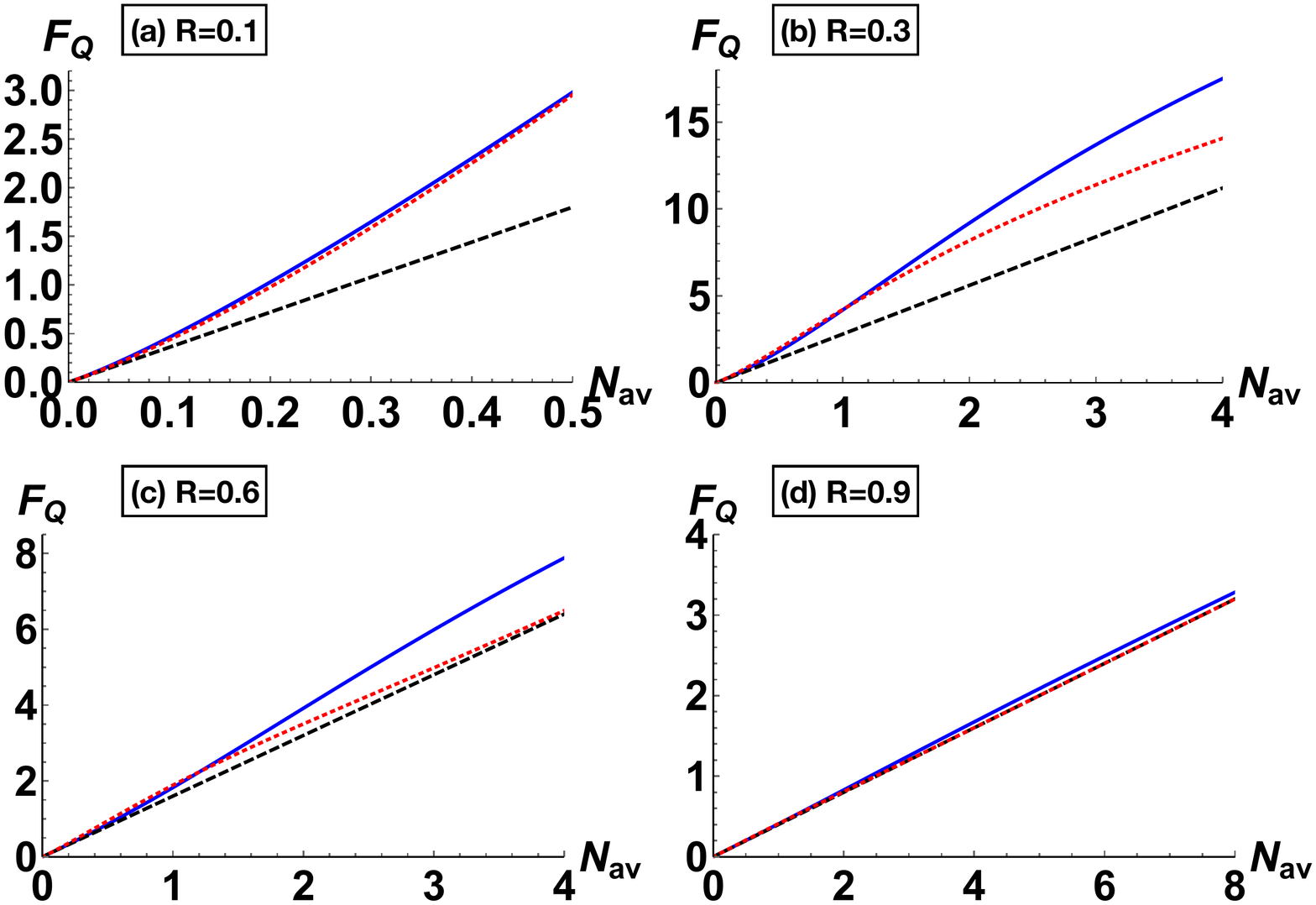}}}
\vspace{-0.25in}
\caption{(Photon loss in one arm) Comparison between the state $|\alpha\rangle_a|\beta\rangle_b+|\beta\rangle_a|\alpha\rangle_b$ and a separable coherent state $|\alpha\rangle_a|\alpha\rangle_b$ for the quantum Fisher information with $N_{\text{av}}=\langle\hat{n}_a\rangle$, in the constraint of input mean photon number. $R$ is the photon loss rate.
Black dashed line ($|\alpha\rangle_a|\alpha\rangle_b$), red dotted curve ($|\alpha\rangle_a|0\rangle_b+|0\rangle_a|\alpha\rangle_b$), 
and blue solid curve ($|\alpha\rangle_a|\beta\rangle_b+|\beta\rangle_a|\alpha\rangle_b$).
(a) $\beta=-0.1\alpha$, (b) $\beta=0.2\alpha$, (c) $\beta=0.4\alpha$, and (d) $\beta=0.5\alpha$.
}
\label{fig:fig7}
\end{figure}

\section{Output state entanglement}
In quantum metrology, the best strategy is to achieve the highest QFI as we can. In the absence of photon loss, the more the degree of entanglement, the higher the QFI. However, in the presence of photon loss, a state in a high degree of entanglement can be faster to lose the amount of entanglement than a state in a low degree of entanglement with the increasing photon loss rate $R$. This is explained by analyzing the entanglement of the output optimal entangled coherent states. For a measure of entanglement in the output state, we consider the negativity which is described with the absolute value of the sum of the negative eigenvalues of a partial transposed state \cite{VW02}, 
$E_N(\rho)\equiv\frac{||\rho^{T_a}||_1-1}{2}=|\sum_i\mu_i|$, where $||\rho^{T_a}||_1$ is the trace norm and $\mu_i$ are the negative eigenvalues of a partial transposed state $\rho^{T_a}$. 

\subsection{Equal losses in both arms}
The negativity of the output state is given by
\begin{eqnarray}
E_N(\rho_{out})=\frac{e^{(1-R)(\beta-\alpha)^2}-1}{2(1+e^{(\beta-\alpha)^2})}.
\end{eqnarray}
Given a fixed value of $\alpha$, we show that a high entangled state is faster to lose entanglement than a low entangled state with increasing $R$, as shown in Fig. 8.
At $R=0$, the amount of entanglement represents their initial entanglement. Then, the output state entanglement is exhibited with a nonzero photon loss rate. At some points of $R$, the initial low entangled state can have relatively more entanglement than the initial high entangled state. The surviving output entanglement contributes to produce relatively high quantum Fisher information by its interference terms. In Fig. 8 (d), the amount of entanglement is distinctively reversed with the increasing photon loss rate.
%After photon loss, the initial low entangled state can have relatively more entanglement than the initial high entangled state. It is also achieved under the constraint of the input mean photon number. The survived entanglement contributed to produce relatively high QFI.  
Thus, in the presence of photon loss, it is valuable to prepare a less entangled coherent state initially with the increasing photon loss rate. 

\begin{figure}
\centerline{\scalebox{0.31}{\includegraphics[angle=0]{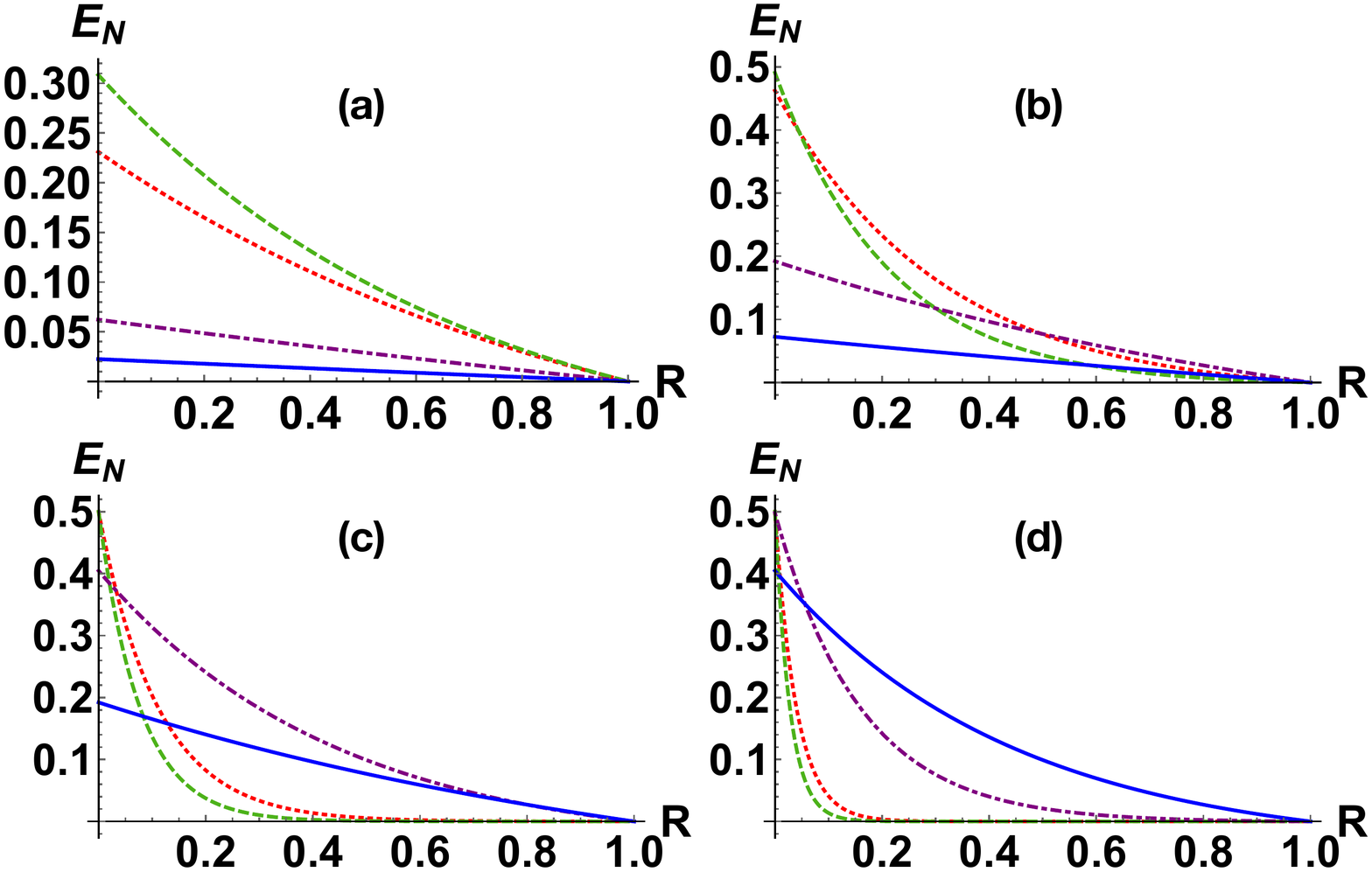}}}
\vspace{-0.5in}
\caption{(Photon loss in both arms) Output entanglement of the optimal entangled coherent state $|\alpha\rangle_a|\beta\rangle_b+|\beta\rangle_a|\alpha\rangle_b$ as a function of  photon loss rate $R$ for fixed values of (a) $\alpha=1$, (b) $\alpha=1.8$, (c) $\alpha=3$, and (d) $\alpha=5$.
Green dashed curve ($\beta=-0.2\alpha$),  red dotted curve ($\beta=0$),  purple dotdashed curve ($\beta=0.5\alpha$), and 
blue solid curve ($\beta=0.7\alpha$).
}
\label{fig:fig8}
\end{figure}

\subsection{Losses in one arm}
The negativity of the output state is given by
\begin{eqnarray}
E_N(\rho_{out})=\bigg|\frac{B_1-\sqrt{B^2_2-4B_3}}{16N_T}\bigg|,
\end{eqnarray}
where $B_1=8(1-e^{-R(\beta-\alpha)^2/2})(1-e^{-(1+T)(\beta-\alpha)^2/2})$,
$B_2=8(1-e^{-R(\beta-\alpha)^2/2})(e^{-T(\beta-\alpha)^2/2}-e^{-(\beta-\alpha)^2/2})$,
$B_3=16(1+e^{-R(\beta-\alpha)^2/2})^2(1-e^{-T(\beta-\alpha)^2})(1-e^{-(\beta-\alpha)^2})$, 
and $N_T=2(1+e^{-(\beta-\alpha)^2})$.
This shows a similar behavior to Fig. 8. For a fixed value of $\alpha$,  a high entangled state is faster to lose entanglement than a low entangled state with increasing $R$, as shown in Fig. 9.  After photon loss, the initial low entangled state can have relatively more entanglement than the initial high entangled state. 

\begin{figure}
\centerline{\scalebox{0.31}{\includegraphics[angle=0]{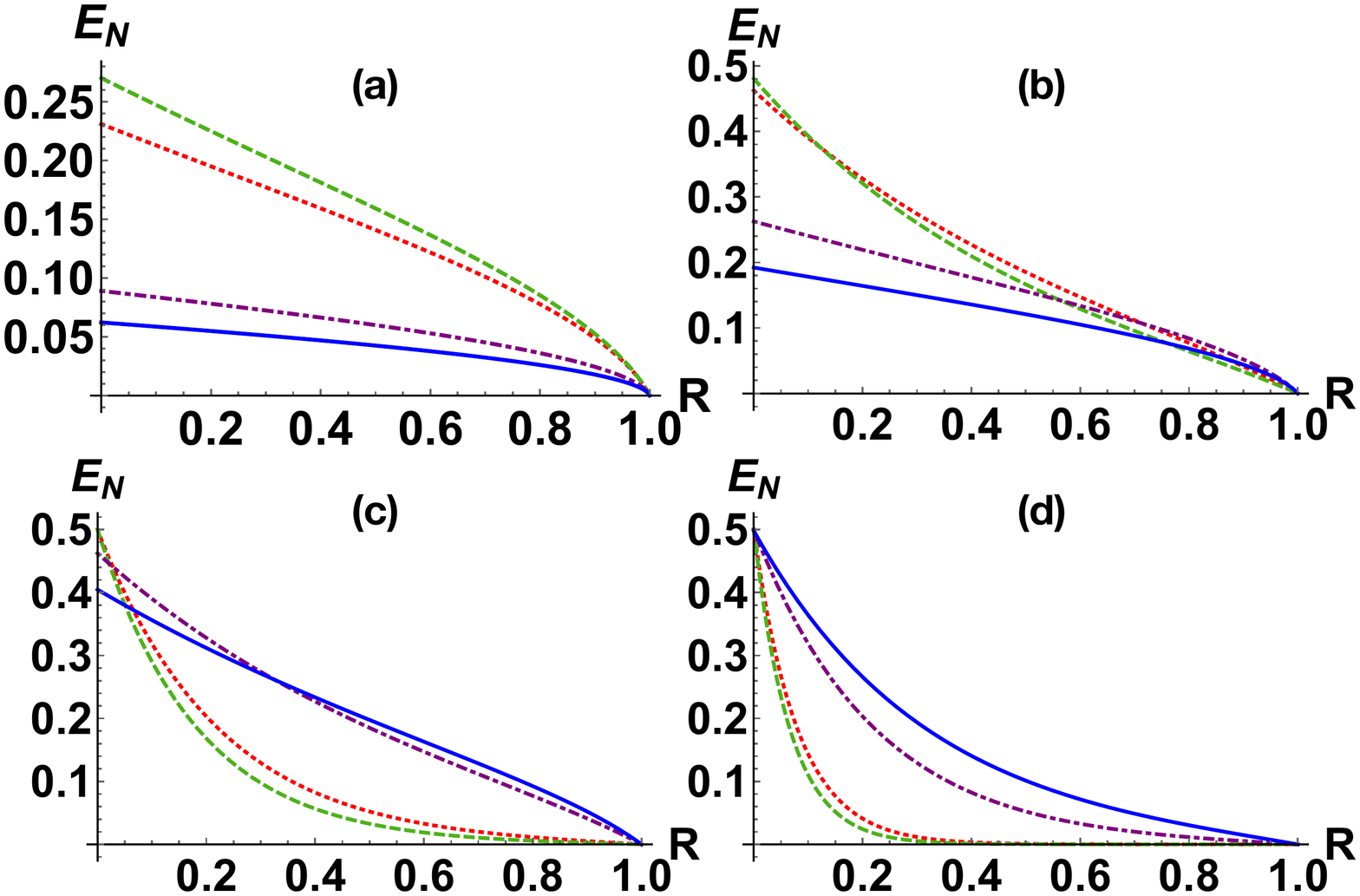}}}
\vspace{-0.4in}
\caption{(Photon loss in one arm) Output entanglement of the optimal entangled coherent state $|\alpha\rangle_a|\beta\rangle_b+|\beta\rangle_a|\alpha\rangle_b$ as a function of  photon loss rate $R$ for fixed values of (a) $\alpha=1$, (b) $\alpha=1.8$, (c) $\alpha=3$, and (d) $\alpha=5$.
Green dashed curve ($\beta=-0.1\alpha$), red dotted curve ($\beta=0$),  purple dotdashed curve ($\beta=0.4\alpha$), and 
blue solid curve ($\beta=0.5\alpha$).
}
\label{fig:fig9}
\end{figure}

\section{Optimal measurement}
Now we turn to the optimal measurement for the quantum Fisher information.
The corresponding optimal measurement is derived with the symmetric logarithmic derivative, the eigenbasis of which represents the optimal measurement basis \cite{Paris}. 
The SLD is defined as $\hat{L}_{\phi}=2\sum_{n,m}\frac{\langle\lambda_m| \partial_{\phi}\rho_{\phi}|\lambda_n\rangle}{\lambda_n+\lambda_m}|\lambda_m\rangle\langle\lambda_n|$, where $\rho_{\phi}=\sum_n \lambda_n|\lambda_n\rangle\langle\lambda_n|$, 
$\langle \lambda_n|\lambda_m\rangle=\delta_{n,m}$, and 
$\partial_{\phi}\rho_{\phi}=\frac{\partial\rho_{\phi}}{\partial\phi}=\sum_n(\partial_{\phi}\lambda_n|\lambda_n\rangle\langle\lambda_n|+\lambda_n|\partial_{\phi}\lambda_n\rangle\langle\lambda_n|+\lambda_n|\lambda_n\rangle\langle\partial_{\phi}\lambda_n|)$.

\subsection{Equal losses in both arms}
Using the SLD formula, we derive the SLD of the output state $\rho_{\phi}$ as
\begin{eqnarray}
\hat{L}_{\phi}=A(|\lambda_-\rangle\langle \lambda_+|- |\lambda_+\rangle\langle \lambda_-|),
\end{eqnarray}
where 
\begin{eqnarray}
&&A=\frac{iT(\alpha^2-\beta^2)(e^{-T(\beta-\alpha)^2}+e^{-R(\beta-\alpha)^2})}{(1+e^{-(\beta-\alpha)^2})\sqrt{1-e^{-2T(\beta-\alpha)^2}}},\\
&&|\lambda_{\pm}\rangle=\frac{(|\sqrt{T}\alpha e^{i\phi}\rangle_a|\sqrt{T}\beta\rangle_b\pm |\sqrt{T}\beta e^{i\phi}\rangle_a|\sqrt{T}\alpha\rangle_b)}{\sqrt{2(1\pm e^{-T(\beta-\alpha)^2})}}. \nonumber
\end{eqnarray}
One of the corresponding eigenbases is $|\lambda_+\rangle \pm i|\lambda_-\rangle$. 
To obtain the QFI, we need to perform a measurement with the eigenbasis $|\lambda_+\rangle \pm i|\lambda_-\rangle$
which consists of $|\sqrt{T}\alpha e^{\phi}\rangle_a|\sqrt{T}\beta\rangle_b$ and $|\sqrt{T}\beta e^{\phi}\rangle_a|\sqrt{T}\alpha\rangle_b$. This requires a correlated detection setup which performs measurements either on the state of $|\sqrt{T}\alpha e^{\phi}\rangle_a|\sqrt{T}\beta\rangle_b$ or on the state of 
$|\sqrt{T}\beta e^{\phi}\rangle_a|\sqrt{T}\alpha\rangle_b$. Each detection setup can be implemented with heterodyne (also called double homodyne) detection on each output mode. However, it is not known how to correlate the different detection setups, and we could not find it either. We leave it for our future work.
Thus, the optimal measurement basis of the output state is a correlated measurement basis which is not achieved with currently known detection schemes.

\subsection{Losses in one arm}
For a photon loss only in one arm of the interferometry, the SLD formula is also given by Eq. (11) but all the components are derived as
\begin{eqnarray}
&&A=\frac{-iT(\alpha^2-\beta^2)(e^{-\frac{1}{2}(1+T)(\beta-\alpha)^2}+e^{-\frac{R}{2}(\beta-\alpha)^2})}
{(1+e^{-(\beta-\alpha)^2})\sqrt{1-e^{-(1+T)(\beta-\alpha)^2}}},\nonumber\\
&&|\lambda_{\pm}\rangle=\frac{(|\sqrt{T}\beta e^{i\phi}\rangle_a|\alpha\rangle_b
\pm |\sqrt{T}\alpha e^{i\phi}\rangle_a|\beta\rangle_b)}{\sqrt{2(1\pm e^{-\frac{1}{2}(1+T)(\beta-\alpha)^2})}}.
\end{eqnarray}
We also need to perform a correlated measurement
which consists of the states $|\sqrt{T}\alpha e^{\phi}\rangle_a|\beta\rangle_b$ and $|\sqrt{T}\beta e^{\phi}\rangle_a|\alpha\rangle_b$. This requires a correlated detection setup which performs measurements either on the state of $|\sqrt{T}\alpha e^{\phi}\rangle_a|\beta\rangle_b$ or on the state of 
$|\sqrt{T}\beta e^{\phi}\rangle_a|\alpha\rangle_b$.

 With a feasible measurement setup, we wonder if the classical Fisher information using photon number resolving detection (PNRD) can approach the quantum Fisher information bound in the presence of photon loss. 
The CFI is given by $F(\phi)=\sum_{n_a,n_b}\frac{1}{P(n_a,n_b|\phi)}[\frac{\partial P(n_a,n_b|\phi)}{\partial \phi}]^2$, where $P(n_a,n_b|\phi)$ is a probability of detecting $n_a$ photons on mode $a$ and $n_b$ photons on mode $b$ for a given phase $\phi$.
An example of the entangled coherent states, i.e., $|\alpha\rangle_a|0\rangle_b+|0\rangle_a|\alpha\rangle_b$, cannot attain the QFI bound by the CFI using the PNRD \cite{LLLN16}. The CFI of the state depends on a phase parameter.
It is known that the quantum Fisher information of a single parameter is independent of the parameter \cite{Helstrom}.
Thus, in the presence of photon loss, we cannot attain the ultimate precision limit of the QFI with the CFI using the PNRD.
For the other feasible measurement setup, we may also consider Gaussian measurement \cite{Gaussian19} that is implemented with general-dyne detection. This will be handled in our next project.

\section{Summary and Discussion}
\begin{figure}
\centerline{\scalebox{0.31}{\includegraphics[angle=0]{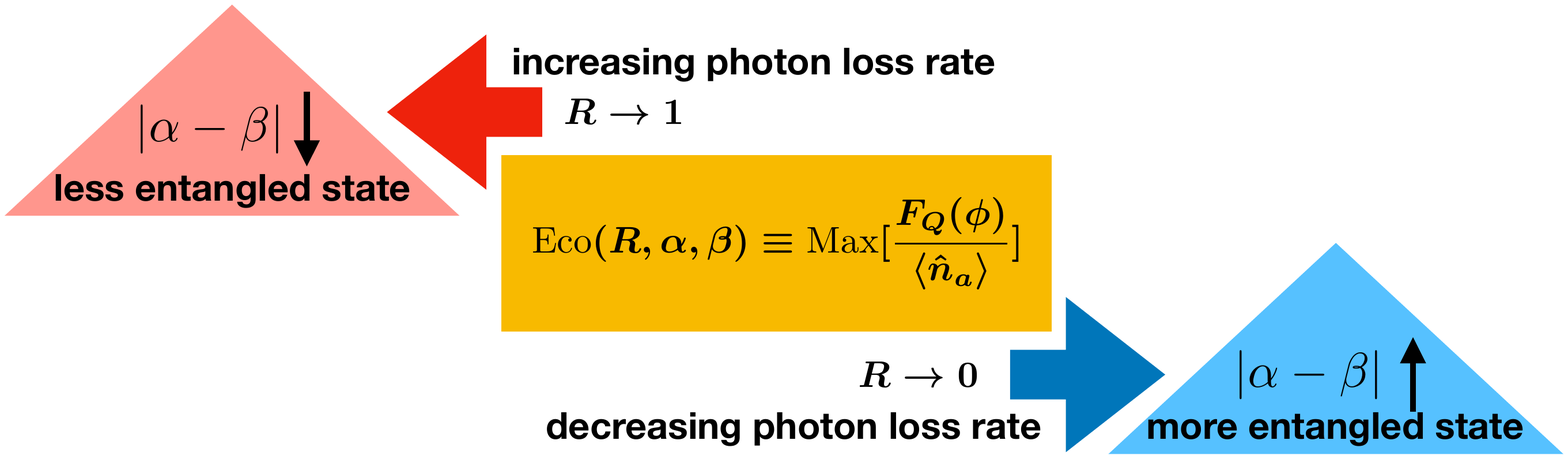}}}
\vspace{-1.6in}
\caption{Economical point as a function $R,~\alpha,$ and $\beta$.
Given a  value of $\alpha$, we find the optimal value of $\beta$ in the range of $-\alpha\leq \beta< \alpha$ to maximize $F_Q/\langle \hat{n}_a\rangle$.
At $R\rightarrow 0$, the optimal value of $\beta$ goes to $-\alpha$ for small $\alpha$ and it becomes zero for large $\alpha$.
}
\label{fig:fig10}
\end{figure}

We investigated an optimal distance of two components ($\alpha$ and $\beta$) in an entangled coherent state 
$|\alpha\rangle_a|\beta\rangle_b+|\beta\rangle_a|\alpha\rangle_b$ for quantum phase estimation in lossy interferometry. Defining the economical point as the quantum Fisher information over the mean photon number of the input mode $a$, we found that the optimal distance of the two components gets smaller with photon loss rate $R$. This represents that initially we need to prepare a less entangled coherent state with increasing $R$. 
We verified that the initial low entangled state can have relatively more entanglement than the initial high entangled state with increasing $R$.
At low photon loss rate ($R\rightarrow 0$), the optimal entangled state is close to
$|\alpha\rangle_a|-\alpha\rangle_b+|-\alpha\rangle_a|\alpha\rangle_b$ for small $\alpha$ and 
$|\alpha\rangle_a|0\rangle_b+|0\rangle_a|\alpha\rangle_b$ for large $\alpha$.
This is summarized in Fig. 10.
We also showed that the optimal entangled coherent state preserves quantum advantage in the presence of photon loss, while surpassing the SIL of a separable coherent state $|\alpha\rangle_a|\alpha\rangle_b$.
Under a fixed input mean photon number, the optimal entangled coherent state is more resilient to photon loss than the separable coherent state, even in high photon loss rates. 
Then we derived the corresponding optimal measurement which is not a simple detection scheme but requires correlation measurement bases. 

It is natural to consider the other type of entangled coherent states, i.e., $|\alpha\rangle_a|\beta\rangle_b-|\beta\rangle_a|\alpha\rangle_b$. 
Since the state does not contain vacuum probability, it is less energy-efficient than $|\alpha\rangle_a|\beta\rangle_b+|\beta\rangle_a|\alpha\rangle_b$ in the condition of the economical point.
Moreover the state can exhibit even worse performance than the separable coherent state $|\alpha\rangle_a|\alpha\rangle_b$ under the constraint of the input mean photon number, as shown in appendix A. That is why we only considered the state $|\alpha\rangle_a|\beta\rangle_b+|\beta\rangle_a|\alpha\rangle_b$ in the lossy quantum-enhanced metrology.

As a further work, we may consider the optimal entangled coherent states propagating in a turbulent atmosphere \cite{SV09,Bohmann17}.
The idea can be also exploited with microwave-optical photon pairs \cite{Stefano15}.

\begin{acknowledgments}
This work was supported by a grant to Quantum Frequency Conversion Project funded by Defense Acquisition Program Administration and Agency for Defense Development.
\end{acknowledgments}

\appendix
\section{The other type of entangled coherent states: $|\alpha\rangle_a|\beta\rangle_b-|\beta\rangle_a|\alpha\rangle_b$}

We assume that $\alpha$ and $\beta$ are real variables. 
The degree of entanglement for the entangled coherent state $|\alpha\rangle_a|\beta\rangle_b-|\beta\rangle_a|\alpha\rangle_b$ is constant as $1$, regardless of $|\alpha-\beta|$. Since it does not contain the probability of the state $|0\rangle_a|0\rangle_b$, the lowest value of the mean photon number is $1/2$. 

\begin{figure}
\vspace{0.1in}
\centerline{\scalebox{0.31}{\includegraphics[angle=0]{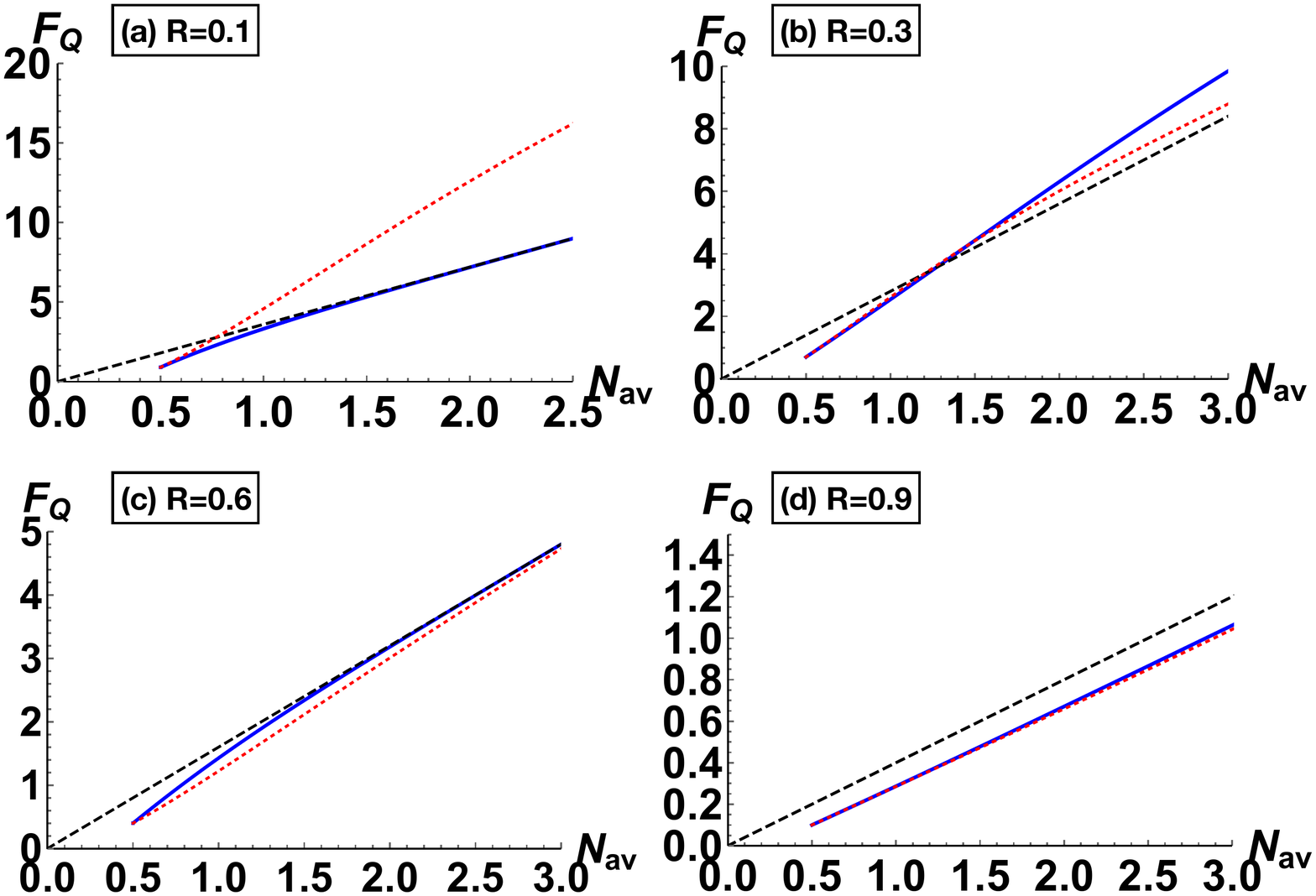}}}
\vspace{-0.3in}
\caption{(Photon loss in both arms) Comparison between the state $|\alpha\rangle_a|\beta\rangle_b-|\beta\rangle_a|\alpha\rangle_b$ and a separable coherent state $|\alpha\rangle_a|\alpha\rangle_b$ for the quantum Fisher information with $N_{\text{av}}=\langle\hat{n}_a\rangle$, in the constraint of input mean photon number. $R$ is the photon loss rate.
Black dashed line ($|\alpha\rangle_a|0\rangle_b$), red dotted curve ($|\alpha\rangle_a|0\rangle_b-|0\rangle_a|\alpha\rangle_b$), 
and blue solid curve ($|\alpha\rangle_a|\beta\rangle_b-|\beta\rangle_a|\alpha\rangle_b$).
(a) $\beta=-\alpha$, (b) $\beta=0.3\alpha$, (c) $\beta=-\alpha$, and (d) $\beta=0.3\alpha$.
}
\label{fig:fig11}
\end{figure}

\begin{figure}
\vspace{0.1in}
\centerline{\scalebox{0.31}{\includegraphics[angle=0]{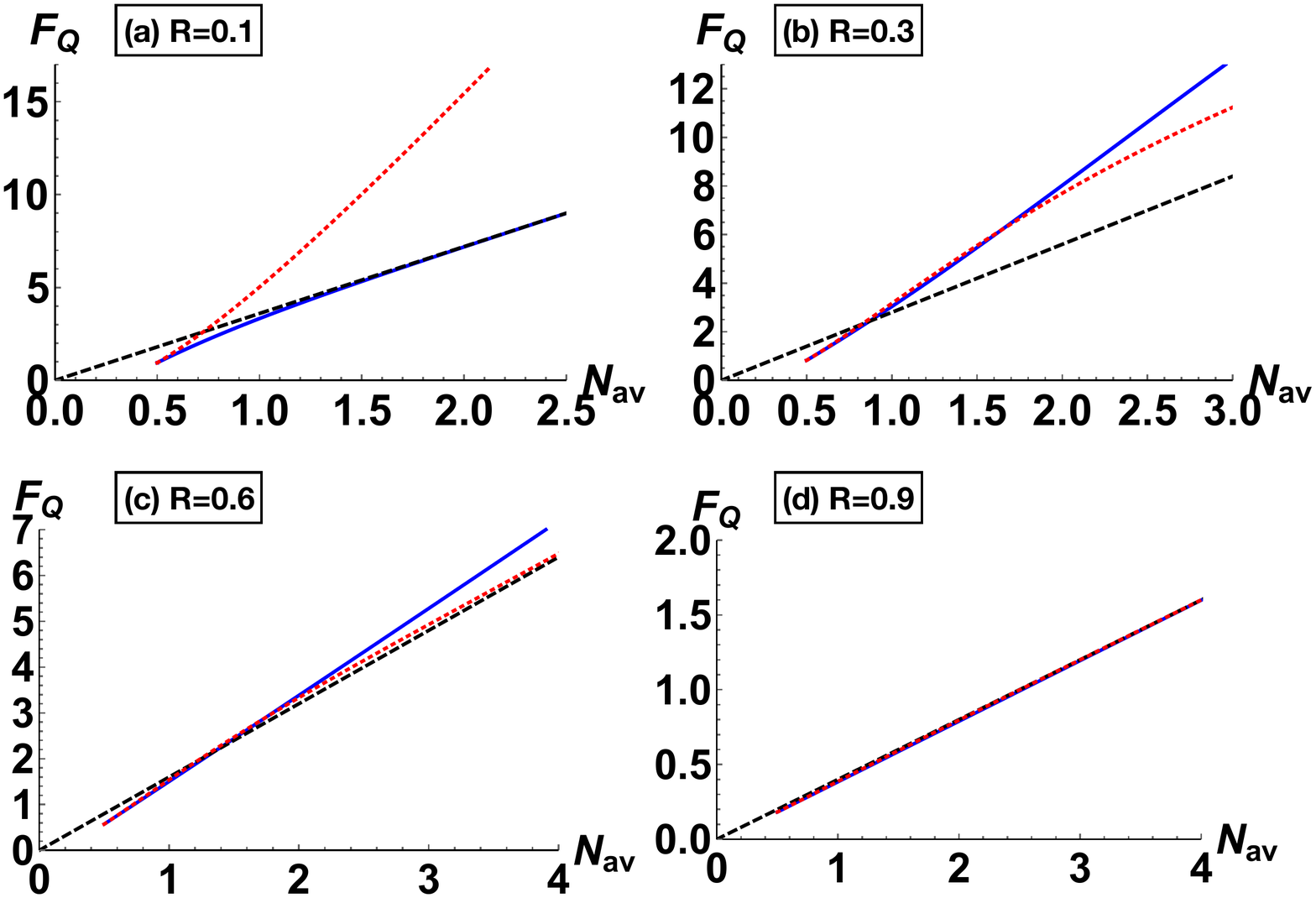}}}
\vspace{-0.3in}
\caption{(Photon loss in one arm) Comparison between the state $|\alpha\rangle_a|\beta\rangle_b-|\beta\rangle_a|\alpha\rangle_b$ and a separable coherent state $|\alpha\rangle_a|\alpha\rangle_b$ for the quantum Fisher information with $N_{\text{av}}=\langle\hat{n}_a\rangle$, in the constraint of input mean photon number. $R$ is the photon loss rate.
Black dashed line ($|\alpha\rangle_a|0\rangle_b$), red dotted curve ($|\alpha\rangle_a|0\rangle_b-|0\rangle_a|\alpha\rangle_b$), 
and blue solid curve ($|\alpha\rangle_a|\beta\rangle_b-|\beta\rangle_a|\alpha\rangle_b$).
(a) $\beta=-\alpha$, (b) $\beta=0.3\alpha$, (c) $\beta=0.6\alpha$, and (d) $\beta=0.4\alpha$.
}
\label{fig:fig12}
\end{figure}

After experiencing a phase shifting operation and photon loss in both arms, the output state is given by
%\begin{widetext}
\begin{eqnarray}
\rho_{\phi}&=&\frac{1}{2[1-e^{-(\beta-\alpha)^2}]}\nonumber\\
&\times&\bigg[ |\sqrt{T}\alpha e^{i\phi}\rangle_a\langle \sqrt{T}\alpha e^{i\phi}|\otimes |\sqrt{T}\beta \rangle_b\langle \sqrt{T}\beta |\nonumber\\
&&+ |\sqrt{T}\beta e^{i\phi}\rangle_a\langle \sqrt{T}\beta e^{i\phi}|\otimes |\sqrt{T}\alpha \rangle_b\langle \sqrt{T}\alpha | \nonumber\\
&& -e^{-R(\beta-\alpha)^2}\bigg( |\sqrt{T}\alpha e^{i\phi}\rangle_a\langle \sqrt{T}\beta e^{i\phi}|\otimes |\sqrt{T}\beta \rangle_b\langle \sqrt{T}\alpha|\nonumber\\
&&+ |\sqrt{T}\beta e^{i\phi}\rangle_a\langle \sqrt{T}\alpha e^{i\phi}|\otimes |\sqrt{T}\alpha \rangle_b\langle \sqrt{T}\beta|\bigg) \bigg],
\end{eqnarray}
%\end{widetext}
where $T=1-R$. Then we derive the quantum Fisher information with the eigenvalues and eigenvectors of the output state. 
In Fig. 11, we compare the QFI of the entangled coherent state $|\alpha\rangle_a|\beta\rangle_b-|\beta\rangle_a|\alpha\rangle_b$ with that of a separable coherent state $|\alpha\rangle_a|\alpha\rangle_b$ under a fixed input mean photon number. 
For small $R$, the coherent state outperforms the entangled coherent state in the small input mean photon number.
For large $R$, the coherent state outperforms the entangled coherent state in the whole region of the input mean photon number.

Assuming a photon loss only in the mode $a$ of Fig. 1, we derive the output state as
%\begin{widetext}
\begin{eqnarray}
\rho_{\phi}&=&\frac{1}{2[1-e^{-(\beta-\alpha)^2}]}
\bigg[ |\sqrt{T}\alpha e^{i\phi}\rangle_a\langle \sqrt{T}\alpha e^{i\phi}|\otimes |\beta \rangle_b\langle \beta |\nonumber\\
&&+ |\sqrt{T}\beta e^{i\phi}\rangle_a\langle \sqrt{T}\beta e^{i\phi}|\otimes |\alpha \rangle_b\langle \alpha | \nonumber\\
&& -e^{-R(\beta-\alpha)^2}\bigg( |\sqrt{T}\alpha e^{i\phi}\rangle_a\langle \sqrt{T}\beta e^{i\phi}|\otimes |\beta \rangle_b\langle \alpha|\nonumber\\
&&+ |\sqrt{T}\beta e^{i\phi}\rangle_a\langle \sqrt{T}\alpha e^{i\phi}|\otimes |\alpha \rangle_b\langle \beta|\bigg) \bigg],
\end{eqnarray}
%\end{widetext}
where $T=1-R$. Using the corresponding quantum Fisher information, we compare the state $|\alpha\rangle_a|\beta\rangle_b-|\beta\rangle_a|\alpha\rangle_b$ and a separable coherent state $|\alpha\rangle_a|\alpha\rangle_b$ under a fixed input mean photon number, as shown in Fig. 12.
We obtain a similar behavior to the case of the photon loss in both arms, but for large $R$ the entangled coherent state almost overlaps with the coherent state in the QFI curves.

\end{document}